# Cavity Waters Govern Insulin Association and Release: Inferences from Experimental Data and Molecular Dynamics Simulations

Saumyak Mukherjee[a], Sayantan Mondal[a], Ashish Anilrao Deshmukh[b], Balasubramanian Gopal[b] & Biman Bagchi[a,*]

[a]Solid State and Structural Chemistry Unit, Indian Institute of Science, Bangalore-560012, India.
[b]Molecular Biophysics Unit, Indian Institute of Science, Bangalore-560012, India.

**ABSTRACT:** While a monomer of the ubiquitous hormone insulin is the biologically active form in the human body, its hexameric assembly acts as an efficient storage unit. However, the role of water molecules in the structure, stability and dynamics of the insulin hexamer is poorly understood. Here we combine experimental data with molecular dynamics simulations to investigate the shape, structure and stability of an insulin hexamer focusing on the role of water molecules. Both X-Ray analysis and computer simulations show that the core of the hexamer cavity is barrel-shaped, holding, on an average, sixteen water molecules. These encapsulated and constrained molecules impart structural stability to the hexamer. Apart from the electrostatic interactions with $Zn^{2+}$ ions, an intricate hydrogen bond network amongst cavity water and neighboring protein residues stabilizes the hexameric association. These water molecules solvate six glutamate residues inside the cavity decreasing electrostatic repulsions amongst the negatively charged carboxylate groups. They also prevent association between glutamate residues and $Zn^{2+}$ ions and maintain the integrity of the cavity. Simulations reveal that removal of these waters results in a collapse of the cavity. Subsequent analyses also show that the hydrogen bond network among these water molecules and protein residues that face the inner side of the cavity is more rigid with a slower relaxation as compared to that of the bulk solvent. Dynamics of cavity water reveal certain slow water molecules which form the back bone of the stable hydrogen bond network. An efficient modulation of active insulin levels relies on a dynamic equilibrium between the monomer and the hexamer which, in turn, is governed by the relative stability of these two forms (alongside the intermediate dimeric form) under physiological conditions. The analysis presented here suggests a dominant role of structurally conserved water molecules in maintaining the integrity of the hexameric assembly and potentially modulating the dissociation of this assembly into the functional monomeric form.

## INTRODUCTION

Insulin regulates blood glucose levels that influence human health[1-18]. Worldwide it is the primary medication for type-1 diabetes[19-21]. Although, insulin exists in several oligomeric forms[4,22], the monomeric insulin is responsible for its biological activity[23,24]. Insulin monomers, however, are prone to form aggregates either in the body or *in vitro*[25]. An insulin hexamer is the most stable oligomeric state[26,27] and acts as the storage unit of this hormone[12]. The hexamer acquires a stable and symmetric quaternary structure and is stored in the $Zn^{2+}$ rich vesicles of the β-cells of pancreas[11,28-30]. In physiologically optimum conditions, according to the demand of the body, this hexamer breaks into monomers via dimers as an intermediate state[29] leading to a dynamic equilibrium[25] amongst the three forms of insulin. The stability of the insulin hexamer and dimer becomes relevant in this context. The dynamic equilibrium among the oligomeric forms of insulin can be schematically represented as follows,

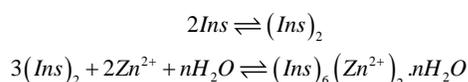

$$2Ins \rightleftharpoons (Ins)_2$$
$$3(Ins)_2 + 2Zn^{2+} + nH_2O \rightleftharpoons (Ins)_6(Zn^{2+})_2 \cdot nH_2O$$

Here 'Ins' stands for insulin monomer. The origin of the increased stability of insulin hexamer is one of the most intriguing problems from a thermodynamic perspective. Not much seems to be known about the detailed molecular level structure and stability of the hexamer, particularly the role of cavity water. This aspect leads to the following interesting questions:

(i) What is the molecular level structure of Insulin hexamer?

(ii) What are the factors responsible for the apparently robust nature of insulin hexamer? That is, what holds it together?

(iii) What role does the confined water molecules play in this context? It should be noted that mobility of these water molecules is related to entropic stabilization of the hexamer.

(iv) Can one rationalize the equilibrium between an insulin monomer, dimer and hexamer? How do the conversions among these quaternary structures take place?

A thermodynamic rationale for these aspects hinges on the structural features of the insulin molecule and its oligomers. An insulin monomer is composed of two chains (A and B) held together by disulphide bonds between Cysteine residues. Some contextual information is presented in supporting information Section S1. Two such monomers combine to form an insulin dimer. In Figure, we schematically depict the dimerization process. The hydrophobic residues Phe-24, Phe-25 and Tyr-26 (from Chain-B) in the anti-parallel β-sheets from two monomers play a key role in dimerization[31]. These residues provide an extended hydrophobic patch (Figure 1) along with π-π stacking of the phenyl rings which largely facilitate this process. Besides this hydrophobic interaction, there are four hydrogen bonds between the back bone atoms involving these residues from each monomer which further drive the process in a forward direction. Karplus and coworkers[32] have shown by MM-GBSA computations that the binding free energy of two insulin monomers to form a dimer is -11.9 kCal mol$^{-1}$. The experimentally determined value[33] of this free energy is -7.2 kCal mol$^{-1}$.

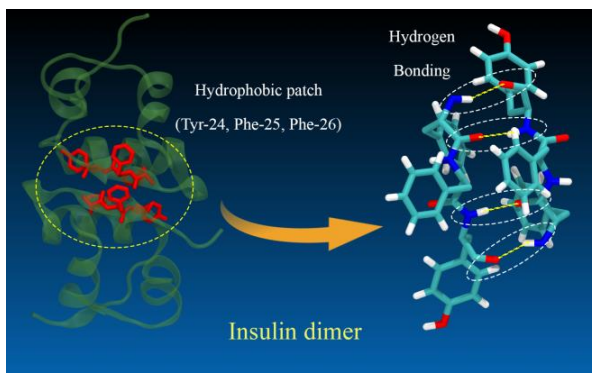

**Figure 1**. Hydrophobic patch at the junction of two monomers in an insulin dimer generated due to the presence of two consecutive Phenylalanine (green) residues (Phe-24, Phe-25) and a Tyrosine (Tyr-26). Four hydrogen bonds exist between them.

Trimerization of these dimers leads to an insulin hexamer in the presence of $Zn^{2+}$ ions. In a $Zn^{2+}$ rich environment, His-10 residues from Chain-B of six monomers get coordinated to two $Zn^{2+}$ ions (3 His-10 coordinate each $Zn^{2+}$) leading to the formation of a hexameric assembly (**Error! Reference source not found.**). A $C_3$ axis of symmetry is present along the straight line joining the two $Zn^{2+}$ ions.

The six Glu-13 residues on the side and two $Zn^{2+}$ ions on top and bottom, along with six His-10 residues define an approximately barrel shaped cavity in the center of the insulin hexamer (Figure 3). This cavity matches nanotube dimensions. Some water molecules are confined in this cavity (Figure 3) and significantly contribute to the stability and firmness of insulin hexamer. The characterization of these water molecules is the prime focus of our study.

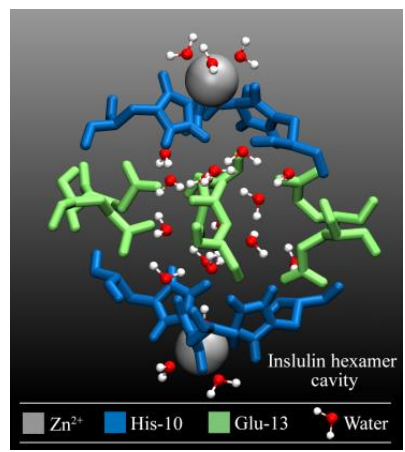

**Figure 3.** Water molecules confined in the insulin hexamer cavity. On an average there are sixteen water molecules inside the barrel shaped cavity out of which six water molecules are perpetually coordinated with two $Zn^{2+}$ ions forming an octahedral geometry along with three His-10 residues.

As mentioned earlier, the hexamer needs to dissociate into monomers to be biologically functional. Though complete mechanism of this dissociation is absent in literature, some studies hint at potential reasons that may trigger this process. Dissociation of insulin hexamer is majorly modulated by the following two contributions: (a) The six Glu-13 residues (each from one monomer) present at the boundary of the cavity partly facilitate this process[4,34]. Repulsion among the negative charges on the Glu-13 side chains causes destabilization of the hexamer leading to dissociation in an environment where $Zn^{2+}$ is scarce. (b) Aspinwall and co-workers have shown that insulin hexamer dissociation is facilitated in alkaline media[35,36]. With the help of amperometric studies, their group has revealed that due to higher extracellular pH insulin secretion rate increases, which is accompanied by dissociation of the hexamer into monomers and subsequent dissolution in blood.

Water in confined systems has always been a field of great interest[37-41]. In the present study we investigate the increased stability and robustness of the insulin hexameric unit focusing on the effects of water molecules confined in the barrel shaped cavity formed at its center, an issue that is still poorly understood.

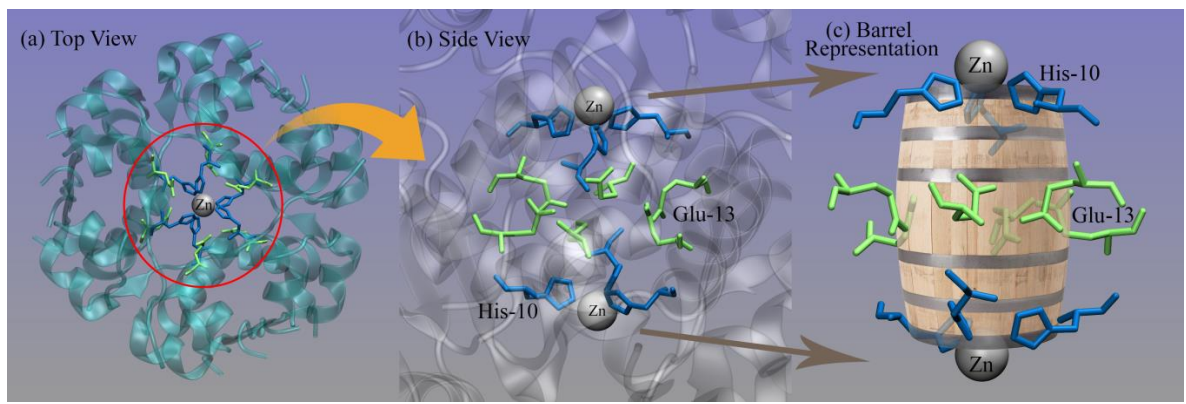

**Figure 2.** Insulin hexamer viewed along the $C_3$ axis (top-view). (b) Two $Zn^{2+}$ ions coordinated by 3 His-10 residues each; boundary of cavity is defined by 1 $Zn^{2+}$ and 3 His-10 residues on top and bottom and 6 Glu-13 residues at the side. (c) Cavity shape in an insulin hexamer is analogous to a barrel.

The organization of the rest of the paper is as follows. In the following section we discuss experimental and theoretical results geared to understand dynamic insulin oligomerization and role of cavity water in stabilization of hexamer. We start the discussion with the energetics of multimerization obtained from quantum chemical computations. Next we discuss observations from x-ray crystallography and study of B-factors of cavity water. This is followed by analysis of conserved water molecules by the superimposition of several previously reported protein structures available in Protein Data Bank (PDB). Thereafter we present results from atomistic molecular dynamics simulations. This includes study of radial distribution function between water molecules and $Zn^{2+}$ ions and hydrogen bond dynamics of cavity water with surrounding protein residues. In the subsequent subsection we show the fate of the cavity in absence of water. Together, these studies identify a causal link between hydration and the hexameric assembly of this hormone.

## RESULTS AND DISCUSSION

### Energetics of multimerization

As mentioned in the previous section, the formation of a dimer from monomers occurs through the association and close contacts of Phe-24, Phe-25 and Tyr-26 residues from two different units. The two major factors governing this process are: (i) The hydrophobic interaction among these residues and (ii) hydrogen bond formation among backbone atoms that form intermolecular and antiparallel β-strands (Figure). In order to obtain the stabilization energy of this association, we choose the aforesaid three residues from each monomer which form anti-parallel β-strands at the junction. Quantum chemical calculations in GAUSSIAN 09[42,43] (details of calculation are given in Computational Section) show that difference between the energies of the associated form and the total energy of the individual strands is -65.97 kcal mol$^{-1}$. This stabilization energy serves as the major driving force for the formation of an insulin dimer. Three such dimers combine to form an insulin hexamer in presence of $Zn^{2+}$ ions with a binding free energy[44] of -26 kCal mole$^{-1}$.

### Conserved water analysis

For analysis of conserved water in insulin hexamer, we have selected 20 insulin structures (from Protein Data Bank) solved by high resolution X-ray crystallography. Details corresponding to each structure are listed in Section S3 (Table S2). We have superimposed these structures in COOT[45] and have calculated their conservation scores (CS) defined as the number of water molecules present at a particular position corresponding to all the twenty structures divided by 20 (expressed as %).

Figure 4(a) shows the CS of water molecules present in the selected PDB structures. Water molecules having CS greater than 70% are shown in green. Most of these highly conserved water molecules are present inside the cavity (dark green) (shown in Figure 4(b) and (c)). However there are certain water molecules having CS > 70% which are not inside the cavity (light green). These molecules sit at the junction of two dimers in the hexameric association, and might play a significant role in holding the two dimers together via bridging H-Bonds. Water molecules bound to $Zn^{2+}$ ions are found to be most conserved (100% CS). These results are consistent with the B-factor values that show that cavity water molecules particularly the ones coordinated to $Zn^{2+}$ ions are most stable. The molecules which are outside the cavity mostly has SC < 70 % (red).

This analysis illustrates that water molecules in the cavity are mostly conserved pointing towards a stable structural modification in the hexamer cavity which becomes crucial when it comes to the stability of the whole hexameric unit.

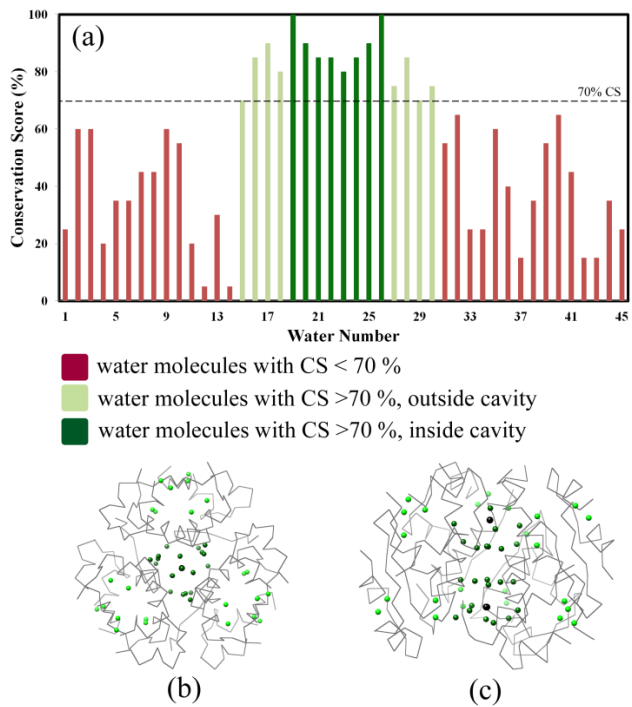

**Figure 4.** (a) Conservation score of water molecules in an insulin hexamer. (b) Top-view and (c) side-view of insulin hexamer with conserved water molecules (cavity: dark green, outside: light green). Water numbering scheme is described in Section S3.

### B-factor of cavity water

B-Factor is a widely used parameter in X-ray crystallography of proteins to get an estimate of the flexibility or mobility of an atom or a part of the system[46]. It is defined as

$$B = 8\pi^2 \langle r^2 \rangle \quad (1)$$

where, $\langle r^2 \rangle$ is the mean square displacement (MSD) of the atom in question. In Figure we present the B-factors of selected water molecules present in the cavity of insulin hexamer from both X-ray crystallography (a) and MD simulation (b). The numbering is according to the position of the molecules starting from one end of the cavity to the other. In this scheme, molecules 1 and 8 are the ones are coordinated to $Zn^{2+}$ ions. These molecules have the least B-factor values. Molecule 5 is situated at a central position of the cavity and has the highest B-factor. Molecules which are hydrogen-bonded to neighboring side chains (Glu, His) have lower B-factors.

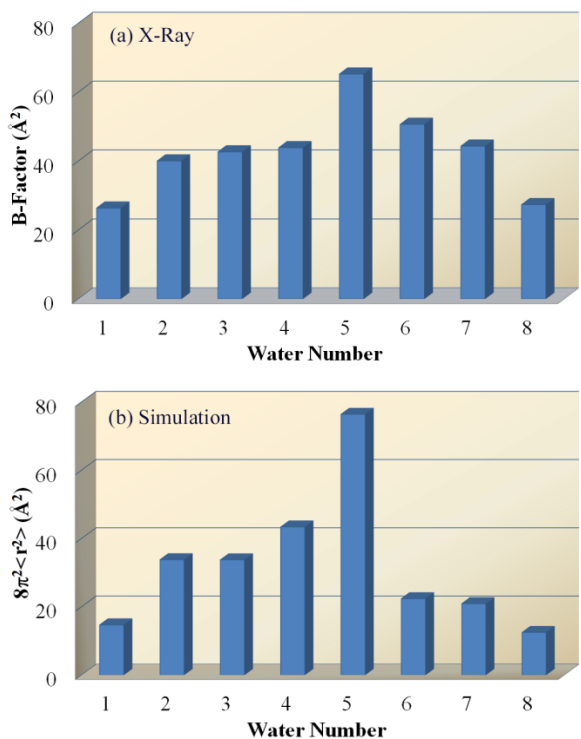

**Figure 5.** B-factors of cavity water. (a) B-factor from X-ray crystallography (b) MSD multiplied by $8\pi^2$ (Equation (1)), obtained from MD simulation.

A lower value of B-factor denotes a more stable system. Therefore, we find that the two ends of the cavity are stable. Water molecules outside the cavity have much higher B-factors, often of the order of ~100 Å$^2$. Hence, cavity water molecules, particularly the ones which are coordinated to $Zn^{2+}$ and hydrogen bonded to amino acid side chains are more ordered than those outside. This provides a firm backbone stabilizing the hexameric association.

### Selection and characterization of cavity water

In computational analyses, exact selection of cavity water is a non-trivial task. Here, we have considered a water molecule to bear cavity properties if it lies within 1.5 nm radius from both the $Zn^{2+}$ ions. Average distance between the two $Zn^{2+}$ ions is 1.3 nm. The extra 0.2 nm distance has been taken to accommodate the water molecules, which are coordinated to $Zn^{2+}$ ions and reside at the two openings of the cavity. A representation of this scheme is given in Figure 6.

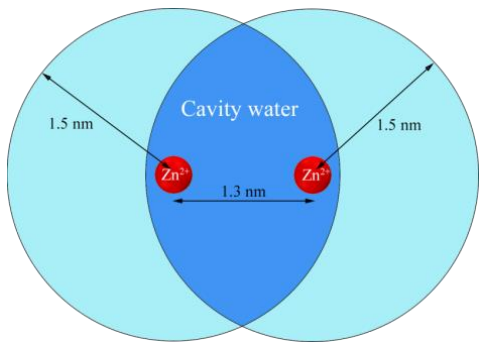

**Figure 6.** Schematic representation of the method of selection of cavity water in computer experiments.

Following this method of characterization of cavity water, average number of molecules is found to be 37. However, core of the cavity between the two $Zn^{2+}$ ions contain 10 water molecules on an average. 6 molecules are coordinated to the two ions. Therefore, these 16 water molecules are present at the heart of the cavity, whereas the remaining 21 molecules remain in close vicinity.

Cavity water differs significantly from bulk water, both in structural and dynamical perspectives. In the present context, we are interested to look into the contribution of these water molecules in stabilizing the hexameric association of insulin. Pair correlation function (or radial distribution function, g(r)) of $Zn^{2+}$ and water molecules shows a sharp peak at 0.2 nm (Figure 7).

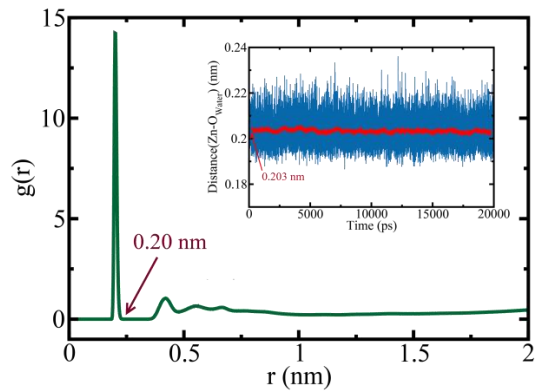

**Figure 7.** Radial distribution function between $Zn^{2+}$ ions and O atoms of water molecules. The sharp peak at 0.2 nm defines the first hydration layer of $Zn^{2+}$ ions constituted by six coordinated water molecules. Inset shows the distance trajectory of one such water molecule from the corresponding $Zn^{2+}$ ion. The red curve denotes running average over periods of 100 ps for the distance.

This peak corresponds to the six highly conserved water molecules that get coordinated to the two $Zn^{2+}$ ions. These 6 molecules have considerably high residence times around two $Zn^{2+}$ ions as is apparent from distance trajectory of one such molecule from the corresponding $Zn^{2+}$ ion (shown in inset of Figure ). Each $Zn^{2+}$ ion, besides being coordinated by 3 water molecules, is also held by 3 His-10 residues. This gives the assembly an octahedral geometry as depicted in Figure 8.

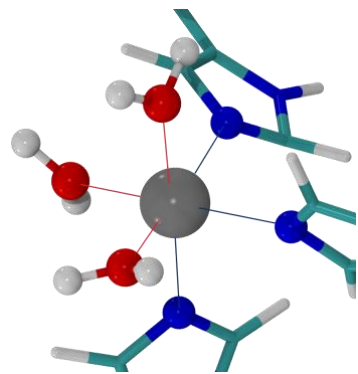

**Figure 8.** Octahedral coordination environment of a $Zn^{2+}$ ion (grey). Red color represents oxygen atoms of water and blue are Nitrogen atoms of His-10 residue.

Such octahedral arrangement is reported in several crystal structures of insulin hexamer in Protein Data Bank (PDB)[3,47,48]. However, some crystal structures report a tetrahedral environment around $Zn^{2+}$ ions which include 3 His-10 residues and one water molecule coordinated to $Zn^{2+}$ ion[49-51]. In simulation, however, we find that the octahedral geometry is conserved throughout the 50 ns trajectory, without the water molecules being exchanged. We performed quantum chemical calculations to estimate the stability of this octahedral complex with respect to hexa-coordinated $Zn^{2+}$-$H_2O$ complex. We compute the relative stability of facial (fac) and meridional (mer) forms of $[Zn(H_2O)_3(Im)_3]^{2+}$ complex with respect to $[Zn(H_2O)_6]^{2+}$ complex. $[Zn(H_2O)_6]^{2+}$ can be considered as a free $Zn^{2+}$ which is yet to enter the cavity. Here, "Im" refers to imidazole which is present in histidine. The results are summarized in Table 1.

**Table 1.** Optimized relative energies of hexa-coordinated $Zn^{2+}$ complexes.

| Molecular formula of complex | Relative stability w.r.t. $[Zn(H2O)_6]^{2+}$ (kcal mol$^{-1}$) |
|---|---|
| fac-$[Zn(H_2O)_3(Im)_3]^{2+}$ | -12.6 |
| mer-$[Zn(H_2O)_3(Im)_3]^{2+}$ | -9.9 |

From the data presented in Table 1, it becomes clear that the facial isomer is more stable than its meridional counterpart. This corroborates with our observation that the complex formed within the insulin cavity indeed has facial geometry. Due to increased stability of $[Zn(H_2O)_3(Im)_3]^{2+}$ complexes with respect to $[Zn(H_2O)_6]^{2+}$, the $Zn^{2+}$ ions get coordinated to the ε-N atom of His-10 residue leading to the formation of insulin hexamer. Thus such coordination is one of the most important factors that stabilize the hexamer unit.

However, further impetus towards the stability is provided by an exceptionally strong hydrogen bond (HB) network formed by the cavity water molecules with the neighboring amino acid side-chains such as glutamate (Glu-13) and histidine (His-10). In order to probe the nature of stability imparted by this HB network we study the HB dynamics of cavity water with the aforesaid protein residues. This is achieved by defining two time correlation functions (TCF) namely intermittent HB TCF ($C(t)$) and continuous HB TCF ($S(t)$) given by the following Equations (2)[52,53].

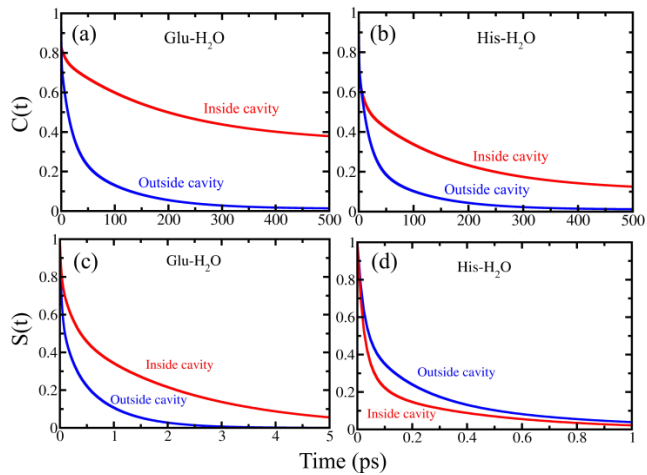

**Figure 9.** HB dynamics of water molecules with nearby protein residues (Glu and His). Red curves depict HB dynamics inside the cavity, whereas the blue ones are for similar pairs outside the cavity. (a) and (b) show intermittent HB TCF, $C(t)$ for Glu and His respectively. (c) and (d) show continuous HB TCF, $S(t)$ in the same sequence.

$$C(t) = \frac{\langle h(0)h(t) \rangle}{\langle h \rangle} \qquad S(t) = \frac{\langle h(0)H(t) \rangle}{\langle h \rangle} \qquad (2)$$

Here $h(t)$ is a population parameter which is defined using a Heaviside step function which attains a value '1' when a particular H-bond exists at time $t$, and '0' otherwise. $H(t)$ is a similar parameter which has a value '1' as long as a H-Bond exists and becomes '0' for the rest of the trajectory when it breaks for the first time. Hence $C(t)$ provides us with the overall structural information of HB network, whereas $S(t)$ estimates its lifetime. Details of theoretical recognition of existence of H-bonds are discussed in Section S4.

In Figure 9, red curves represent HB dynamics for cavity and blue curves represent the same for regions outside the cavity. $C(t)$ for water-Glu and water-His H-Bonds are shown in Figure 9(a) and (b) respectively. In both the cases, relaxation is much slower inside the cavity. It is approximately 3.5 times slower in case of water-Glu H-Bond and 2.5 times slower in case of water-His H-Bond (Table 2). This shows that HB network inside the cavity is much stronger as compared to that outside.

However, the scenario is a little different in case of $S(t)$. Relaxation of this continuous HB TCF is slower for water-Glu H-Bond but faster for water-His H-Bond inside the cavity in contrast to those in the outside. This difference can be attributed to the spatial constrains faced by the coordination of His-10 with $Zn^{2+}$ ions. The side chains of His-10 are not free to move along with the movement of water molecules. This results in recurrent breaking and formation of the corresponding H-Bonds. On the other hand, Glu-13 side chains in the cavity are not constrained, providing the Glu-water H-Bonds a greater life time. This difference reflects the cooperativity between amino acid side chains and water molecules in maintaining the life time of these H-Bonds.

Therefore, combination of these observations indicates that the HB network inside insulin hexamer cavity is extensively mod-ified to form a robust backbone that supports the hexameric association from inside. Hence, H-Bonds among water molecules and surrounding protein residues play a vital role in sustaining the stability and structural features of insulin hexamer.

**Table 2. Average relaxation times for HB dynamics of water and neighboring residues inside and outside the cavity.**

| HB TCF | HB of water with | Average relaxation time(ps) |
|---|---|---|
| C(t) | Glu-13 (*inside*) | 141 |
| | Glu-4,17,21 (*outside*) | 40 |
| | His-10 (*inside*) | 82 |
| | His-5 (*outside*) | 33 |
| S(t) | Glu-13 (*inside*) | 1.240 |
| | Glu-4,17,21 (*outside*) | 0.350 |
| | His-10 (*inside*) | 0.120 |
| | His-5 (*outside*) | 0.184 |

The water molecules which participate in hydrogen bonding and coordination have very high residence times (residence time distribution of cavity water with respect to bulk is shown in Figure S3 of Section S5). These molecules exhibit slower translational and rotational dynamics (Sections S6 and S7). Relaxation of total dipole moment correlation function is also much slower in cavity as compared to bulk water (Section S8). Slower dynamics of the cavity water ensures a robust structure at the interior of insulin hexamer. Experimental evidence of structurally stable water distribution in the cavity is also provided by investigation of electron density map. Details of this study are provided in the Section S9. We find that molecules which are hydrogen bonded to neighboring polar residues in the cavity have sharp distributions of electron densities, whereas other water molecules possess broader distributions. Furthermore, the peaks in the density of states of cavity water are blue shifted with respect to bulk (Figure S9) denoting a more structured cavity interior as compared to the bulk. This once again brings forward the unique nature of structural stability of water in insulin hexamer cavity which stands out as a major support to the insulin hexameric association.

### Fate of cavity in absence of water

The analyses presented thus far seem to suggest that cavity water is significantly responsible for maintenance of stability of insulin hexamer. In order to further substantiate this observation we monitor the stability of the cavity in absence of water in computer simulation. (It is to be noted that the initial structure of this simulation has no water molecules in the cavity; however water is present outside the cavity.) Analysis of the trajectory reveals that when water is removed, the cavity breaks down within a few femtoseconds.

This collapse is found to follow the following sequential steps. The $Zn^{2+}$ ions come closer to an average distance of 0.8 nm (as opposed to a separation of 1.3 nm in presence of water). In the absence of water molecules, coordination sites on $Zn^{2+}$ become vacant and Glu-13 with its negatively charged side-chain is seen to coordinate with the $Zn^{2+}$ ions. This helps in reducing the repulsive interactions among the negatively charged carboxylate moieties. We observe that two Glu-13 residues get coordinated to each $Zn^{2+}$ ion thereby altering the $Zn^{2+}$-His-10 coordination (Figure ). These choreographed steps lead to a collapse of the cavity disrupting the symmetric arrangement of the insulin hexamer into a less ordered aggregate. The role of the conserved water molecules in initiating this dissociation suggests a dominant role of hydration forces in the hexameric insulin assembly.

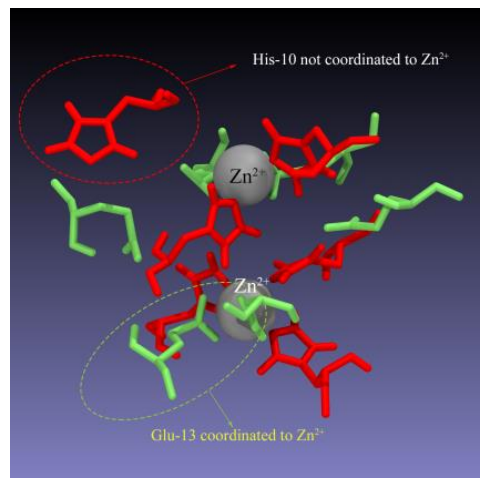

**Figure 10.** Fate of the hexamer cavity in absence of water. The two $Zn^{2+}$ ions come closer and Glu-13 gets coordinated to them. His-10 coordination with $Zn^{2+}$ is disturbed.

### ▪ CONCLUSION

Despite the pivotal role played by the insulin hexamer as the storage of insulin in human body, it was surprising that the role of cavity water in its stability had not been examined thus far. Such information is clearly required to understand the dynamic equilibrium between the hexamer and the dimer of insulin that partly controls the response of human body to glucose level[25,29].

The most significant observation of this study is that a few water molecules in the core of the hexamer cavity dictate the structural stability of hexamer. These core waters have a distinct signature- low B-factors as inferred from crystallographic data, slow dynamics in molecular dynamics simulations and strong coordination with $Zn^{2+}$ ions, alongside histidine residues. While it was widely assumed that these interactions are responsible for the stability of the insulin hexamer, the crucial role of these conserved waters in screening the electrostatic field due to the carboxylate groups of six Glu-13 residues in the vicinity of hexameric cavity was less understood. Indeed, removal of these waters causes an increase in repulsion among the negative charges of Glu-13 side chains thus destabilizing the hexameric assembly- the cavity collapses within a few ps as Glu-13 perturbs the coordination environment of the $Zn^{2+}$ ions. Hydration is thus the most dominant and yet under-recognized factor that governs insulin aggregation and release.

### ▪ EXPERIMENTAL SECTION

Insulin was obtained from Sigma Aldrich and initial crystallization trials were performed using commercially available crystallization screen from Hampton research. This hormone was crystallized (0.1 M Sodium acetate tri-hydrate, pH 4.6, 2.0 M Sodium chloride) at 7 mg mL$^{-1}$ insulin containing trace amount of zinc chloride. Crystals were obtained and soaked with 10% ethylene glycol in mother liquor prior to data collection. Data was collected at home source and processed by iMOSFLM[54] and scaled using SCALA[55]. The phase information was obtained by molecular replacement method using insulin model (PDB code: 3W7Y). The model was refined using REFMAC5[56] and the fit of the model to the electron density was evaluated using COOT[37]. Data collection and refinement statistics for insulin hexamer and PDB validation statistics are presented in Section S2 of supporting information.

## ■ COMPUTATIONAL SECTION

### Quantum chemical calculations

We use density functional theory (DFT) calculations to enquire the origin of the stability of insulin dimer and hexamer in terms of energetics. First, we calculate the stabilization energy of dimer formation by concentrating on the junction of two monomers. We fix the conformation of that region using the information available from protein data bank (PDB: 3W7Y) and perform single point energy calculation for that domain (highlighted in Figure). B3LYP[57] functional and 6-311G+(d,p) basis set are used for energy calculation. The extra stabilization energy is calculated by subtracting the energy of two monomeric strands from that of the dimeric strand. Further, we probe the stability of hexamer by considering the complex formation among $Zn^{2+}$ ions, imidazole nitrogens (from His-10 residues) and water molecules. We calculate the energies of individual histidine, water and $Zn^{2+}$ along with the energies of other possible complexes. The geometry of the complex here is fixed according to the average equilibrated structure of insulin hexamer in water as obtained from simulation. Basis set superposition error (BSSE) is calculated using counterpoise[58] for every structure. All quantum calculations are performed using Gaussian 09[42,43] package.

### Molecular dynamics simulations

Atomistic molecular dynamics (MD) simulations were performed using GROMACS-4.5.6 package[59] which is a widely accepted and highly efficient MD engine. The initial configuration of the system has been taken from crystal structure available in Protein Data Bank (PDB Code: 3W7Y). The asymmetric unit in 3W7Y has been processed to get the biological unit that is hexamer, using UCSF Chimera-1.11.2[60]. For simulation, we have used GROMOS96 53a6[61] force field for protein and Extended Simple Point Charge Model (SPC/E)[62] for water. Periodic boundary conditions were implemented using a cubic box of 10 nm dimensions with 31620 water molecules in the system.

The total system was energy minimized by a succession of steepest descent and conjugate gradient algorithms. Thereafter the solvent (water) was equilibrated in NPT conditions (T = 300 K and P=1 bar) restraining the positions of protein atoms for 5 ns followed by a similar equilibration under NVT condition (T=300 K). Then the system was subjected to a further 10 ns NPT equilibration (T=300 K and P = 1 bar) without any position restrains. The final production run was carried out in an NVT environment at a temperature of 300 K for 55 ns. The last 50 ns were taken for analyses. Data was dumped at a frequency of 0.1 ps for analyzing static properties and 4 fs for dynamic properties. The equations of motion were integrated using Leap-Frog algorithm with a time step of 1 fs.

In order to maintain a fixed average temperature and pressure, we use Nose-Hoover thermostat[63] ($\tau_t$ = 0.1 ps and two coupling groups, namely protein and non-protein) and Parrinello-Rahman barostat[64] ($\tau_p$=2.0 ps) respectively. A cut-off radius of 10 Å was set for neighbor searching and calculation of non-bonded interactions and all bonds were constrained using LINCS algorithm[65]. For calculation of electrostatic interactions, Particle Mesh Ewald method[66] was used with FFT grid spacing of 1.6 Å.

## ■ ASSOCIATED CONTENT

**Supporting Information.** Structure of insulin monomer, data collection and refinement statistics for X-ray crystallography, details of PDB structures used for conserved water analysis, theoretical definition of hydrogen bond, residence time distribution of cavity water, dynamical features of cavity water compared to bulk, electron density profiles of cavity water, density of states of cavity water compared to bulk.

## ■ AUTHOR INFORMATION

### Corresponding Author

* E-mail: profbiman@gmail.com

## ■ ACKNOWLEDGMENT

We thank Professor Peter Wolynes, Rice University for several interesting discussions. We thank Professor Rajib Biswas for helping in designing some of the figures. We thank the Department of Science and Technology (DST, India) for partial support of this work. B. Bagchi thanks Sir J. C. Bose fellowship for partial support. S. Mondal and A. Deshmukh thank UGC, India for providing research fellowship. S. Mukherjee thanks DST, India for providing INSPIRE fellowship.

## ■ REFERENCES

# Supporting Information

In the main manuscript we discuss the role of water trapped in the cavity at the center of insulin hexamer in stabilization of the hexameric association. By virtue of detailed experimental and computational analyses we establish that apart from the stabilization imparted by strong electrostatic interactions between $Zn^{2+}$ ions and 6 insulin residues (His-10), water molecules centered in the cavity also play a pivotal role in this regard. In this supplementary material, we provide some more information regarding our studies which include details about insulin monomer structure (**Section S1**), data collection and refinement statistics of insulin hexamer for x-ray crystallography (**Section S2**), details of the 20 PDB structures used for conserved water analysis and water numbering scheme in the same (**Section S3**), theoretical definition of hydrogen bond (**Section S4**), residence time distribution of cavity water (**Section S5**), dynamical features of cavity water compared to bulk (including translation, rotation and dielectric relaxation obtained from computer simulations) (**Sections S6, S7 & S8**), electron density profiles of cavity water (**Section S9**) and finally comparison of density of states of cavity water and bulk.

## S1. INSULIN MONOMER

The amino acid sequence in insulin obtained from different organisms is almost the same. The monomer consists of two chains, A and B. Chain-A contains 21 residues whereas chain-B contains 30 residues and are connected by 2 disulphide bonds between Cys-7 (Chain-A) – Cys-7 (Chain-B) and Cys-20 (Chain-A) – Cys-19 (Chain-B). Additionally there is another disulphide bond within Chain-A between Cys-6 and Cys-11. These bonds are of extreme significance in stabilizing the structure of insulin.

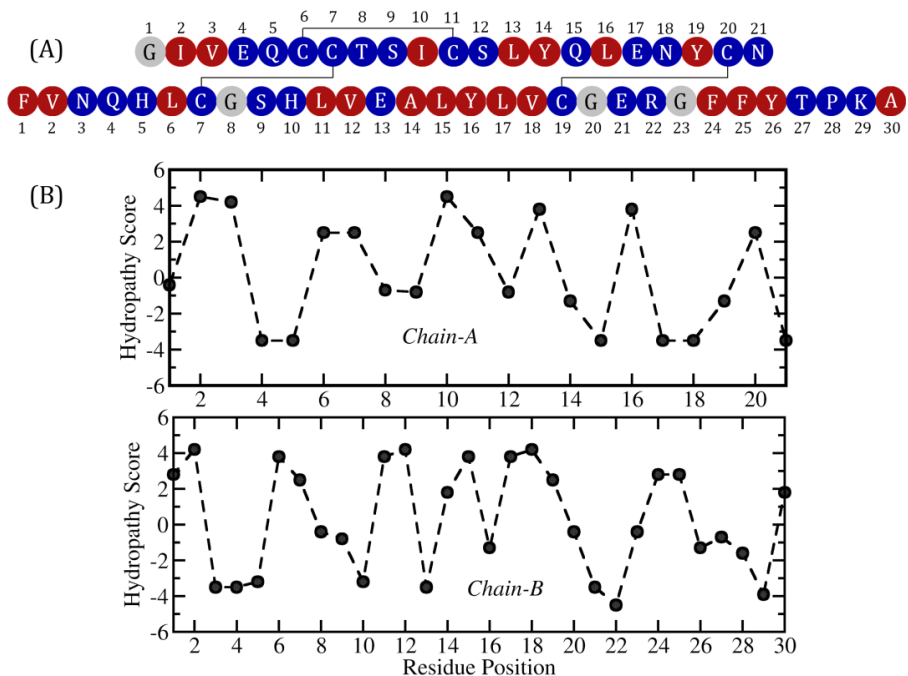

**Figure S1: (A) Amino acid sequence in Insulin. Red denotes hydrophobic residues and blue denotes hydrophilic residues. Neutral glycine residues are color coded with grey. (B) Kyte-Doolittle plot showing the extent of hydrophobicity in Insulin. A more hydrophobic residue has a more positive hydropathy score.**

**Figure S1(A)** shows the color-coded amino acid sequence in the two chains (A and B) of an insulin monomer. Hydrophobic residues are shown in red whereas the hydrophilic residues are in blue. Neutral glycine residues are shown in grey. **Figure S1(B)** is the Kyte-Doolittle hydrophobicity plot, generated according to the value of hydropathy index given in reference[67]. The hydropathy score varies between -4.5 and 4.5 with positive values denoting greater hydrophobicity. The two consecutive hydrophobic residues (phenyl alanine) at 24 and 25 (chain-B) are primarily responsible for the formation of insulin dimer.

## S2.    DATA COLLECTION STATISTICS

**Table S3: Data collection and refinement statistics of Insulin hexamer**

| Data collection statistics | |
|---|---|
| Wavelength (Å) | 1.5418 |
| Resolution (Å) | 39.85-1.85 (39.85-1.85) |
| Unit-cell parameters | a=79.70Å, b= 79.70 Å, c= 36.92 Å<br>α= β=90°, γ =120 |
| Space group | H3 |
| Total No. of reflections | 35989 (2168) |
| No. of unique reflections | 7479 (469) |
| Wilson B factor ($Å^2$) | 34.3 |
| Completeness (%) | 100 (100) |
| Anomalous completeness (%) | 100 (100) |
| Multiplicity | 4.8 (4.6) |
| *$R_{sym}$ (%) | 5.0 (49.2) |
| Mean I/SigI | 12.7 (2.0) |
| CC(1/2) | 0.999 (0.879) |
| **Refinement and model statistics** | |
| †R factor (%) | 17.7 |
| ‡$R_{free}$ (%) | 22.3 |
| RMS Bond lengths (Å) | 0.0188 |
| RMS Bond angles (°) | 1.9433 |
| Ramchandran Favored (%) | 97.80 |
| Ramchandran Allowed (%) | 2.20 |
| Ramchandran outliers | 0 |
| Average B Factor ($Å^2$) | 38.0 |

Values for the outer shell are given in parentheses.

$^*R_{sym} = \dfrac{\sum_{hkl}\sum_{i}|I_i(hkl) - \langle I(hkl)\rangle|}{\sum_{hkl}\sum_{i}I_i(hkl)}$, where $I_i(hkl)$ is the intensity of the $i^{th}$ reflection and $\langle I(hkl)\rangle$ is the average intensity.

$^\dagger R\ factor = \dfrac{\sum_{hkl}|F_{obs} - F_{cal}|}{\sum_{hkl}F_{obs}}$, where $F_{cal}$ and $F_{obs}$ are the calculated and observed structure-factor amplitudes respectively.

$^\ddagger R_{free}$ is calculated like *R factor* but for 5.0% of the total reflections chosen at random and omitted from refinement.

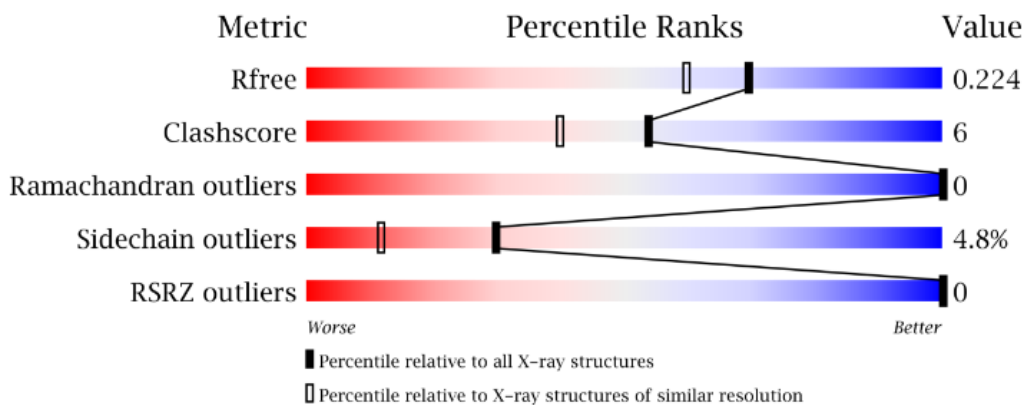

Figure S2: PDB validation statistics of Insulin hexamer.

## S3. CONSERVED WATER ANALYSIS

Details of Insulin Structures taken from PDB

**Table S4: Details of Insulin structure taken from protein data bank**

| Sr. No. | PDB ID | Resolution (Å) | Solvent content (%) | Author determined biological unit |
|---------|--------|----------------|---------------------|-----------------------------------|
| 1 | 3W7Y | 0.92 | 33.39 | Hexamer |
| 2 | 1MSO | 1.00 | 33.22 | Dodecamer |
| 3 | 1EV3 | 1.78 | 39.41 | Dodecamer |
| 4 | 1JCA | 2.5 | 39.43 | Hexamer |
| 5 | 1MPJ | 2.3 | 39.88 | Dimer |
| 6 | 1Q4V | 2.0 | 39.48 | Hexamer |
| 7 | 1TRZ | 1.6 | 39.45 | Dimer |
| 8 | 1TYL | 1.9 | 40.45 | Dimer |
| 9 | 2R34 | 2.25 | 38.81 | Dodecamer |
| 10 | 2VJZ | 1.8 | 37.00 | Dodecamer |
| 11 | 3JSD | 2.5 | 41.27 | Dodecamer |
| 12 | 3KQ6 | 1.9 | 32.46 | Dodecamer |
| 13 | 3MTH | 1.9 | 38.09 | Dimer |
| 14 | 4GBC | 1.78 | 33.71 | Dodecamer |
| 15 | 4GBK | 2.4 | 33.94 | Dodecamer |
| 16 | 5CO6 | 1.8 | 34.02 | Dodecamer |
| 17 | 5HPU | 2.2 | 40.57 | Dodecamer |
| 18 | 2INS | 2.5 | 36.80 | Dodecamer |
| 19 | 3P2X | 2.0 | 34.11 | Hexamer |
| 20 | 4INS | 1.5 | 36.02 | Dimer |

Water Numbering Scheme

**Figure 5** shows the conservation scores of water molecules present in and around insulin hexamer cavity as obtained from 20 crystal structures (asymmetric units) from Protein Data Bank. The water molecules having conservation scores (CS) > 70 % occupy the central position of the graph (light and dark green). The rest have conservation scores < 70 %. The water molecules present inside the cavity (dark green) are from 19 to 26, with the two extremes being coordinated to $Zn^{2+}$. Water molecules having numbers 15 to 18 and 27 to 30 are outside the cavity, but possess CS > 70 %. The molecules marked red have CS < 70 % and are numbered as 1 to 14 and 31 to 45.

Water numbers 1 to 8 in **Figure 4** correspond to water molecules 19 to 26 serially in **Figure 5.**

## S4. THEORETICAL REALIZATION OF HYDROGEN BOND

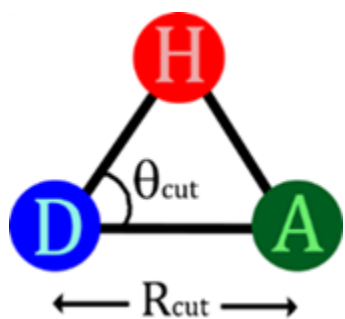

**Figure S3: Geometrical criteria of determining the presence of hydrogen bond. D is the donor atom while A is the acceptor atom. Bond HA is the hydrogen bond.**

The presence of a HB is determined by two geometrical parameters: (a) cut-off distance ($R_{cut}$), determined by first minimum of pair-correlation function and (b) angle cut-off ($\theta_{cut}$), generally taken to be 30° (**Figure S3**)[52].

## S5. RESIDENCE TIME DISTRIBUTION OF CAVITY WATER

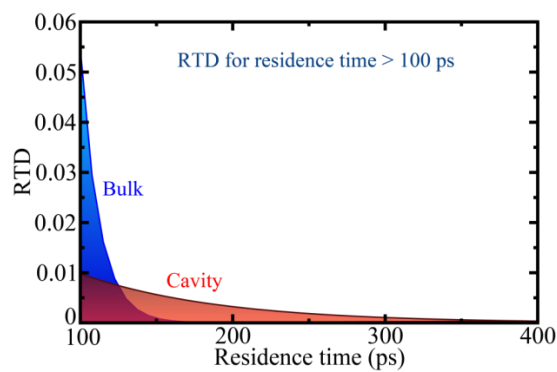

**Figure S4: Residence time distribution of water molecules in insulin hexamer cavity (for RT > 100 ps). Unlike bulk, cavity contains certain water molecules that are much long lived and less mobile.**

Cavity water shows a range of residence times, as presented in the residence time distribution (RTD) in **Figure S4**. Here, RTD is plotted for those water molecules which have RT greater than 100 ps. Insulin cavity comprises of some water molecules which are much slower than bulk. The cavity water selected in our study comprises of two domains: i) core water molecules which are confined within the cavity and ii) water molecules outside the cavity but in close vicinity of the same, so that they are strongly influenced by cavity environment. The first category consists of approximately 16 molecules on an average, which are slower than the rest. The long tail present in the RTD of cavity water accounts for such slow water which are absent in bulk (blue). The distribution for bulk shown here is with respect to a similar cavity created in neat water maintaining the geometrical aspects in the original insulin hexamer cavity.

## S6. TRANSLATIONAL DYNAMICS

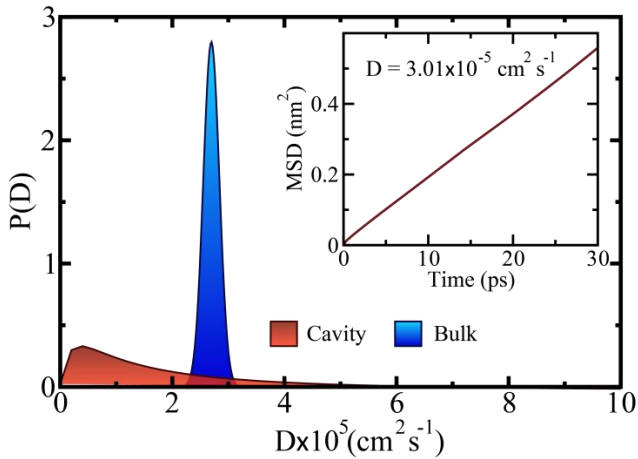

**Figure S5: Distribution of self-diffusion coefficients of cavity water molecules. Inset shows average mean square displacement and diffusion coefficient.**

To understand the nature of translational diversity of the water molecules confined in the cavity of insulin hexamer, we have calculated their self-diffusion coefficients (D) from mean square displacements (MSD) ($\langle r^2 \rangle$) according to Einstein's formula given by **Equation (1)**.

$$\langle r^2 \rangle = 6Dt \qquad (3)$$

We monitored the molecules according to their residence times in the cavity. We performed the calculations of MSD for the duration in which the respective molecule is present in the hexamer cavity. The average value of D for cavity water has been found to be $3.01 \times 10^{-5}$ cm$^2$s$^{-1}$ whereas for SPC/E water the value is $2.7 \times 10^{-5}$ cm$^2$s$^{-1}$. These values suggest that cavity waters are somewhat translationally faster than or almost comparable to the bulk, which is not exactly correct. As mentioned previously, cavity water selected in our study contains some water molecules which do not reside in the core of the cavity. These molecules, being translationally faster than core-water, shift the average to a higher value. However, the distribution of self-diffusion coefficients of cavity water brings out the true heterogeneous picture (**Figure S5**). The distribution which has a *log-normal* nature shows that several water molecules present in the hexamer cavity are much slower than bulk water. These are those molecules which are confined in the core of the cavity.

## S7. ROTATIONAL DYNAMICS

Orientational relaxations of most of these confined water molecules are slower than bulk. **Figure S6** shows the average orientational time correlation functions, $r_1(t)$ and $r_2(t)$ (**Equations (2)** and **(3)**), corresponding to the first and second rank spherical harmonics for cavity and bulk water.

$$r_1(t) = \langle \mathbf{P}_1(\hat{\mu}_0 . \hat{\mu}_t) \rangle; \text{ where } \mathbf{P}_1(x) = x \qquad (4)$$

$$r_2(t) = \langle \mathbf{P}_2(\hat{\mu}_0 . \hat{\mu}_t) \rangle; \text{ where } \mathbf{P}_2(x) = \tfrac{1}{2}(3x^2 - 1) \quad (5)$$

Here, we monitor the orientation of a particular O-H bond vector in each water molecule. $\mathbf{P_1}$ and $\mathbf{P_2}$ are first and second rank Legendre polynomials respectively and $\mu_t$ is the unit vector along the monitored O-H bond at time $t$.

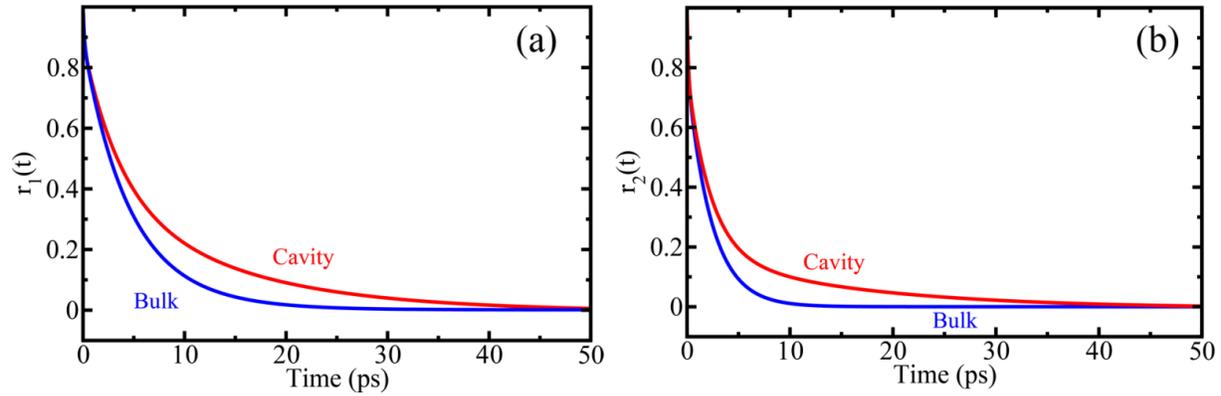

**Figure S6: Average orientational relaxation of cavity water molecules compared to that of bulk. (a) First rank Legendre polynomial. (b) Second rank Legendre polynomial. Relaxation is slower in case of cavity water.**

We have fitted the relaxations to multi-exponential forms according to **Equation (4)**.

$$r(t) = \sum_i a_i e^{-\frac{t}{\tau_i}} \quad (6)$$

The fitting parameters are presented in **Table S5**.

**Table S5: Multi-exponential fitting parameters for orientational correlation**

| Property | Domain | $a_1$ | $\tau_1$ (ps) | $a_2$ | $\tau_2$ (ps) | $a_3$ | $\tau_3$ (ps) | $\langle\tau\rangle$ (ps) |
|---|---|---|---|---|---|---|---|---|
| $r_1(t)$ | Cavity | 0.23 | 61.36 | 0.62 | 5.57 | 0.15 | 0.295 | 17.61 |
|  | Bulk | 0.87 | 4.86 | 0.13 | 0.196 | - | - | 4.25 |
| $r_2(t)$ | Cavity | 0.20 | 14.23 | 0.62 | 2.07 | 0.18 | 0.001 | 4.12 |
|  | Bulk | 0.81 | 2.34 | 0.19 | 0.138 | - | - | 1.92 |

$\langle\tau\rangle$ is the average orientational relaxation time calculated according to **Equation (5)**.

$$\langle\tau\rangle = \int_0^t dt' r(t') = \sum_i a_i \tau_i \quad (7)$$

From **Table S5** it is seen that for first rank correlation, $\langle \tau_1 \rangle_{cav} / \langle \tau_1 \rangle_{bulk} = 4.14$, whereas the ratio for second rank correlation is $\langle \tau_2 \rangle_{cav} / \langle \tau_2 \rangle_{bulk} = 2.14$. The slowest component of $r_1(t)$ for cavity water is almost 12 times greater than its bulk analogue, whereas the same for $r_2(t)$ is ~6. This orientational slowness is caused due to hindered rotation which is a result of H-bonding with protein side-chains and strong electrostatic interaction with $Zn^{2+}$ ions.

The slow water molecules observed in these theoretical analyses correspond to the ones which have high low B-fators. These molecules being less mobile are capable of maintaining a well-structured hydrogen bond network at the core of the cavity, which ultimately results in stabilizing the insulin hexamer structure.

## S8. DIELECTRIC RELAXATION

Dielectric relaxation (DR) also characterizes cavity water to be distinct from bulk. Time correlation function of total dipole moment fluctuation (given by **Equation (6)**) of cavity water follows a triexponential law, the average relaxation time being 14.80 ps, which is ~ 1.7 times slower than bulk (8.86 ps). Cavity DR has a slow component of 25.43 ps which characterizes 54 % of the decay.

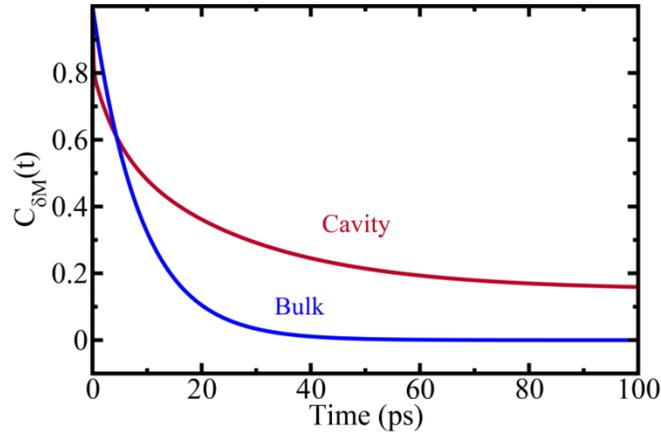

**Figure S7: Total dipole moment fluctuation correlation function of cavity and bulk water. Relaxation is slower in cavity.**

$$C_{\delta M}(t) = \frac{\langle \delta M(0) \delta M(t) \rangle}{\langle \delta M(0) \delta M(0) \rangle} \qquad (8)$$

**Table S6: Fitting parameters of dielectric relaxation.**

| Domain | $a_1$ | $\tau_1$ (ps) | $a_2$ | $\tau_2$ (ps) | $a_3$ | $\tau_3$ (ps) | $\langle \tau \rangle$ (ps) |
|---|---|---|---|---|---|---|---|
| Cavity | 0.54 | 25.43 | 0.24 | 4.34 | 0.22 | 0.122 | 14.80 |
| Bulk | - | - | - | - | - | - | 8.86 |

## S9. ELECTRON DENSITY PROFILES

Electron density map (mF$_o$–DF$_c$) (obtained from difference between observed and calculated structure factor amplitudes) of 2 PDB structures (PDB: 5E7W and 3W7Y) were superimposed in COOT[45]. Details of selected PDB structures are listed in **Table S7**.

**Table S7: Details if selected PDB structures**

| Sr. No. | PDB ID | Resolution (Å) | Solvent content (%) | Number of molecules in asymmetric unit |
|---|---|---|---|---|
| 1 | 5E7W | 0.95 | 35.00 | 2 |
| 2 | 3W7Y | 0.92 | 33.39 | 2 |

Water molecules present in asymmetric unit are considered for studying their electron density shape profile variation. Water molecule nomenclature is given as per PDB 1OS3. Only superimposed water molecules are selected for studying their electron density shape profile. The extent of electron density distribution is calculated in terms of full width half maxima (FWHM) at 1 sigma level of electron density map (mF$_o$–DF$_c$) for selected water molecules in the above mentioned structures. Average value of extent of electron density distribution is considered for each water molecule corresponding to above mentioned PDB structures.

Water molecules (622 (D)) interacting with 13 Glu possesses sharp electron density distribution profile while water molecules (613 (B), 614 (B) and 604 (D)) present in the second solvation shell around Zn$^{2+}$ ions have more divergence in distribution. Water molecule (627 (B)) has sharp electron density distribution profile due to hydrogen bond interactions with neighbouring water molecules (622 (D)). Water molecule (618 (B)) has sharp electron density distribution profile due hydrogen bond interactions with 9 Ser (B). Water molecules in the cavity have sharp distribution profile as compared to the water molecules at entrance of the core. Other than cavity waters (612 (B) and 614 (B)) shows diverged distribution profile due to multiple hydrogen bonding interactions possible with neighboring polar residues. Water molecules (610 (B) and 606 (B)) hydrogen bonded with 10 His residues coordinating Zn$^{2+}$ ions shows sharp distribution profile.

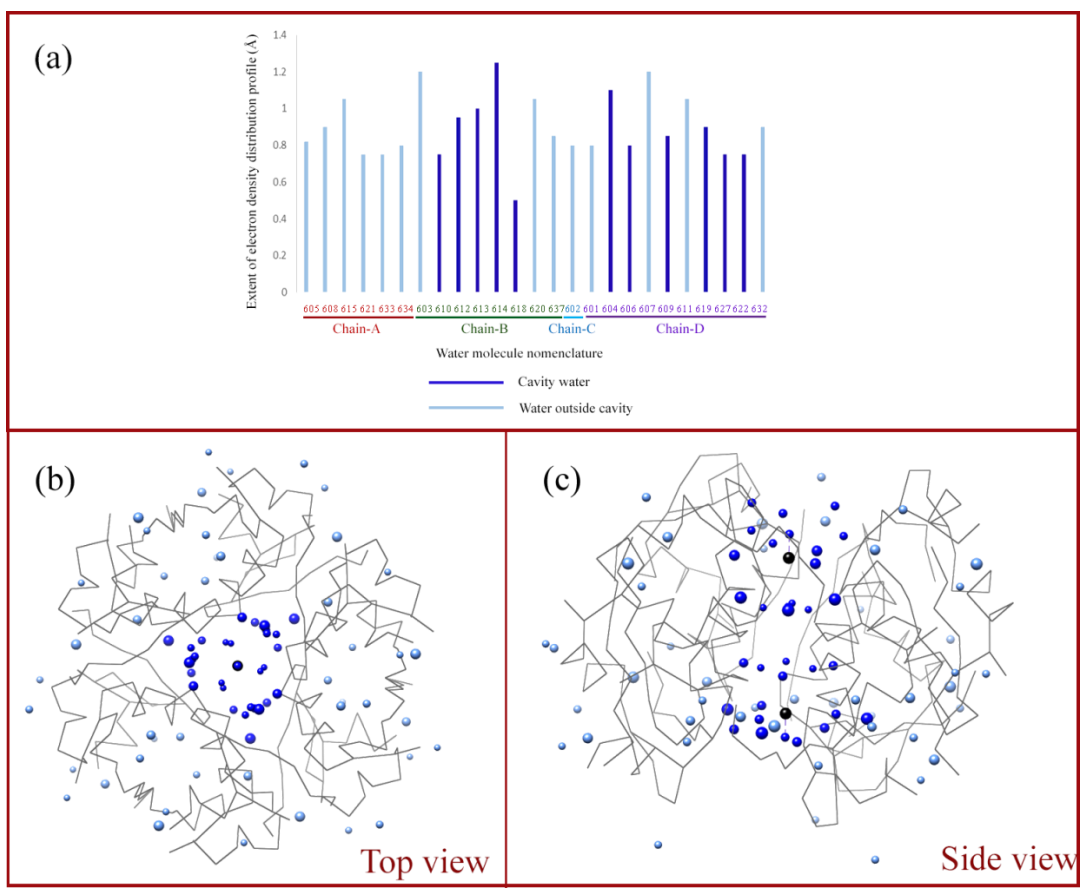

Figure S8: (a) Graphical representation of extent of electron density distribution of water molecules in superimposed Insulin hexamer structures. (b) Top-view and (b) side-view of electron density distribution shape profiles of water molecules in Insulin hexamer. C-alpha chain is shown in grey color. Cavity water molecules are colored dark blue while other water molecules are colored as sky blue. Radii of the spheres represent the width of electron density distributions.

## S10. DENSITY OF STATES

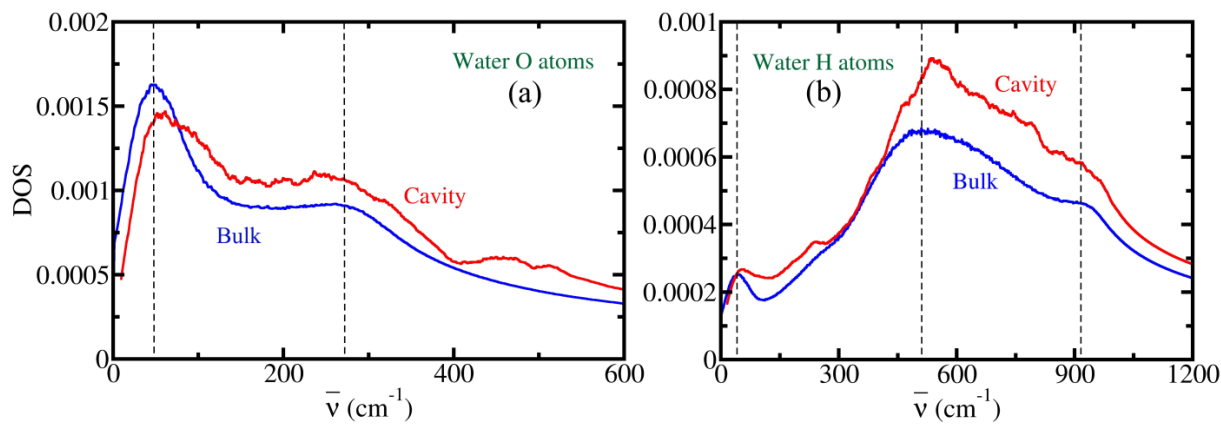

Figure S9: Density of States (DOS) for water molecules within insulin hexamer cavity (red) compared to that of bulk (blue); (a) Oxygen atoms and (b) Hydrogen atoms.

Density of states (DOS) of molecules gives us an idea about the degrees of freedom enjoyed by them. This includes motions like translation, rotation, vibration, libration etc. Inspection of **Figure S9** shows that the cavity water suffers a blue shift of the major peaks in DOS denoting such motions, in cases of both H and O atoms of water. This increase in energy points towards a water network in the cavity which is more structured and robust as compared to that of the bulk.